\documentstyle[12pt]{article}
\begin{document}

\title{{\bf Quantum Cosmology}
\thanks{Alberta-Thy-09-06, hep-th/0610121, in {\em The Future of
Theoretical Physics and Cosmology: Celebrating Stephen Hawking's 60th
Birthday}, edited by G.~W.~Gibbons, E.~P.~S.~Shellard, and S.~J.~Rankin
(Cambridge University Press, Cambridge, 2003), pp. 621-648.}}

\author{
Don N. Page
\thanks{Internet address:
don@phys.ualberta.ca}
\\
Institute for Theoretical Physics\\
Department of Physics, University of Alberta\\
Room 238 CEB, 11322 -- 89 Avenue\\
Edmonton, Alberta, Canada T6G 2G7
}

\date{(2002 Mar. 29)}

\maketitle
\large
\begin{abstract}
\baselineskip 14 pt

	A complete model of the universe needs at least three parts:
(1) a complete set of physical variables and dynamical laws for them,
(2) the correct solution of the dynamical laws, and
(3) the connection with conscious experience.
In quantum cosmology, item (2) is the quantum state of the cosmos.
Hartle and Hawking have made the `no-boundary' proposal,
that the wavefunction of the universe is given by a path integral
over all compact Euclidean 4-dimensional geometries and matter fields
that have the 3-dimensional argument of the wavefunction on their
one and only boundary.  This proposal is incomplete in several ways but
also has had several partial successes, mainly when one takes the
zero-loop approximation of summing over a small number of complex
extrema of the action.  This is illustrated here by the
Friedmann-Robertson-Walker-scalar model.
In particular, new results are discussed when the scalar field has
an exponential potential, which generically leads to an infinite number
of complex extrema among which to choose.

\end{abstract}
\normalsize
\baselineskip 14 pt
\newpage

\section{Motivation for a quantum state of the cosmos}

	A complete model of the universe neads at least three parts:
	
\begin{enumerate}
\item A complete set of physical variables (e.g., the arguments
of the wavefunction) and dynamical laws (e.g., the Schr\"{o}dinger
equation for the wavefunction, the algebra of operators
in the Hilbert space, or the action for a path integral.)
Roughly speaking, these dynamical laws tell how things change with time.
Typically they have the form of differential equations.

\item The correct solution of the dynamical laws
(e.g., the wavefunction of the universe).
This picks out the actual quantum state of the cosmos
from the set of states that would obey the dynamical laws.
Typically a specification of the actual state would involve
initial and/or other boundary conditions for the dynamical laws.

\item The connection with conscious experience
(e.g., the laws of psycho-physical experience)
These might be of the form that tells what conscious experience
occurs for a possible quantum state for the universe,
and to what degree each such experience occurs
(i.e., the measure for each set of conscious experiences
\cite{Page1995,Page2001}).

\end{enumerate}

	Item 1 alone is called by physicists a TOE or `theory of
everything,' but it is not complete by itself. In this chapter I shall
focus on Item 2, but even Items 1 and 2 alone are not complete, since
by themselves they do not logically determine what, if any, conscious
experiences occur in a universe. For example, suppose we have a unique
quantum state for a TOE that consists of some completion of string/M
theory. If this completion is anything like our present partial
knowledge of string/M theory or of other dynamical laws in physics, it
will not by itself answer the question of why most alert humans are
usually consciously aware of their visual sensations but not of their
heartbeats.

	(Of course, one might postulate some principle whereby one
would most of the time be consciously aware of visual sensations but
not of one's heartbeat, but I do not see how such a principle would be
directly derivable from the quantum state and a full set of dynamical
laws that are at all similar to our presently known incomplete
approximations to these laws.  One might propose that awareness of
visual sensations would be more useful in the survival of the fittest
than awareness of one's heartbeat, but it is not obvious to me how
conscious awareness contributes to survival, even though it may be
correlated with some physical information processing that is useful for
survival.  When Fermilab Director Robert Wilson was asked by a
Congressional committee what Fermilab contributes to the defense of the
nation, he is reported to have responded, ``Nothing.  But it helps make
the nation worth defending.''  Similarly, consciousness may contribute
nothing to the survival of an organism, but it may make the organism
worth surviving and may be the selection mechanism that makes us
aware of being conscious organisms.)

	But even before we attack the difficult question of the
relationship between conscious experience and the rest of physics, it
is clear that much of what we try to describe in physics depends not
only on the dynamical laws but also on some features of the quantum
state of the cosmos \cite{Har}. For example, the observation that the
universe has far less entropy than one might imagine, so that the
entropy tends to increase (the second law of thermodynamics), cannot be
purely a consequence of the familiar types of dynamical laws but must
also depend on the quantum state of the cosmos.

\section{The Hartle-Hawking proposal for the quantum state}

	Here I shall focus on Item 2, the quantum state of the cosmos,
and in particular focus on a proposal by Hawking \cite{H,HNP} and by
Hartle and Hawking \cite{HH} for this quantum state. They have proposed
that the quantum state of the universe, described in canonical quantum
gravity by what we now call the Hartle-Hawking wavefunction, is given
by a path integral over compact four-dimensional Euclidean geometries
and matter fields that each have no boundary other than the
three-dimensional geometry and matter field configuration that is the
argument of the wavefunction.

	(Thus this proposal is sometimes called the `no-boundary'
proposal. However, just as in this volume my main Ph.D. advisor Kip
Thorne has rebelled against his advisor John Wheeler in relabeling
Wheeler's `no-hair' conjecture as the `two-hair' conjecture, to count
the mass and angular momentum of a Kerr black hole, so I am enboldened
to rebel half as much against my other Ph.D. advisor, Stephen Hawking,
by relabeling his `no-boundary' proposal as the `one-boundary'
proposal, to count the one boundary of the path integral that is the
argument of the wavefunction.)

	In particular, the wavefunction for a three-geometry given by a
three-metric $g_{ij}(x^k)$, and for a matter field configuration
schematically denoted by $\phi^A(x^k)$, where the three-metric and the
matter field configuration are functions of the three spatial
coordinates $x^k$ (with lower-case Latin letters ranging over the three
values $\{1,2,3\}$), is given by the wavefunction
\begin{equation}
\psi[g_{ij}(x^k),\phi^A(x^k)]
= \int{\mathcal{D}}[g_{\mu\nu}(x^{\alpha})]
{\mathcal{D}}[\phi^{\Omega}(x^{\alpha})]
e^{-I[g_{\mu\nu},\phi^{\Omega}]},
\end{equation}
where the path integral is over all compact Euclidean
four-dimensional geometries that have the three-dimensional
configuration $[g_{ij}(x^k),\phi^A(x^k)]$ on their one and only
boundary.  Here a four-geometry are given by a four-metric
$g_{\mu\nu}(x^{\alpha})$, and four-dimensional matter field histories
are schematically denoted by $\phi^{\Omega}(x^{\alpha})$, both functions
of the four Euclidean spacetime coordinates $x^{\alpha}$ (with
lower-case Greek letters ranging over the four values $\{0,1,2,3\}$).

	The Hartle-Hawking `one-boundary' proposal is incomplete in
various ways.  For example, in quantum general relativity, using the
Einstein-Hilbert-matter action, the path integral is ultraviolet
divergent and nonrenormalizable \cite{GS}. This nonrenormalizability
also occurs for quantum supergravity \cite{Des}. String/M theory gives
the hope of being a finite theory of quantum gravity (at least for each
term of a perturbation series, though the series itself is apparently
only an asymptotic series that is not convergent.)  However, in
string/M theory it is not clear what the class of paths should be in
the path integral that would be analogous to the path integral over
compact four-dimensional Euclidean geometries without extra boundaries
that the Hartle-Hawking proposal gives when general relativity is
quantized.

	Another way in which the Hartle-Hawking `one-boundary' proposal
is incomplete is that conformal modes make the Einstein-Hilbert action
unbounded below, so the path integral seems infinite even without the
ultraviolet divergence \cite{GHP}. If the analogue of histories in
string/M theory that can be well approximated by low-curvature
geometries have actions that are similar to their general-relativistic
approximations, then the string/M theory action would also be unbounded
below and apparently exhibit the same infrared divergences as the
Einstein-Hilbert action for general relativity.  There might be a
uniquely preferred way to get a finite answer by a suitable restriction
of the path integral, but it is not yet clear what that might be.

	A third technical problem with the Hartle-Hawking path integral
is that one is supposed to sum over all four-dimensional geometries,
but the sum over topologies is not computable, since there is no
algorithm for deciding whether two four-dimensional manifolds have the
same topology.  This might conceivably be a problem that it more
amenable in string/M theory, since it seems to allow generalizations of
manifolds, such as orbifolds, and the generalizations may be easier to
sum over than the topologies of manifolds.

	A fourth problem that is likely to plague any proposal for the
quantum state of the cosmos is that even if the path integral could be
uniquely defined in a computable way, it would in practice be very
difficult to compute.  Thus one might be able to deduce only certain
approximate features of the universe from such a path integral.

	Despite the difficulties of precisely defining and evaluating
the Hartle-Hawking `one-boundary' proposal for the quantum state of the
universe, it has had a certain amount of partial successes in
calculating certain approximate predictions for highly simplified toy
models:

\begin{enumerate}

\item Lorentzian-signature spacetime can emerge in a WKB limit of an
analytic continuation
\cite{HH,HNP}.

\item The universe can inflate to large size
\cite{HNP}.

\item Models can predict near-critical energy density
\cite{HNP,HP}.

\item Models can predict low anisotropies
\cite{HL}.

\item Inhomogeneities start in ground states and so can fit cosmic
microwave background data
\cite{HawHal}.

\item Entropy starts low and grows with time
\cite{Haw,Page85,HLL}.

\end{enumerate}

\section{Zero-loop quantum cosmology and
FRW-scalar models}

	One can avoid many of the problems of the Hartle-Hawking
path-integral, and achieve some partial successes for the `one-boundary'
proposal, by taking
\begin{equation}
\psi[g_{ij}(x^k),\phi^A(x^k)] \approx \psi_{\rm 0-loop}
 = \sum_{\rm some\ extrema}e^{-I[g_{\mu\nu},\phi^{\Omega}]},
\label{1}
\end{equation}
summing over a small set of extrema of the Euclidean action $I$,
generally complex classical solutions of the field equations.

	Even at this highly simplified approximation to the path
integral, there is the question of which extrema to sum over, since
typically there are infinitely many.

	A simple class of models that has often been considered is the
$k=+1$ Friedmann-Robertson-Walker-scalar model, in which the
three-geometry boundary is an $S^3$ with radius $\sqrt{2G/3\pi}\,a_b$
and the (real) scalar field takes the homogeneous value
$\sqrt{3/4\pi G}\,\phi_b$ on the boundary,
where Newton's constant is $G$
and I have set $\hbar$ and $c$ equal to unity.  (The numerical factors
used to define the physical radius and scalar field value in terms of
rescaled values $a_b$ and $\phi_b$ enable one to dispense with similar
factors in the expressions involving $a_b$ and $\phi_b$.)

	Then the zero-loop approximation gives
\begin{equation}
\psi(a_b,\phi_b) \approx \psi_{\rm 0-loop}(a_b,\phi_b)
 = \sum_{\rm some\ extrema}e^{-I(a_b,\phi_b)},
\label{2}
\end{equation}
where $I(a_b,\phi_b)$ is the Euclidean action of a classical solution
that is compact and has the $S^3$ geometry and homogeneous scalar field
as its one and only boundary.

	`One-boundary' FRW-scalar histories have a time parameter $t$
that can be taken to run from 0 (at a regular `center') to 1
(at the boundary), and then to have $\phi = \phi(t)$ and four-metric
 \begin{equation}
 ds^2 = 
 \left({2G\over 3\pi}\right)\left[N^2(t)dt^2 + a^2(t)d\Omega_3^2\right],
 \label{3}
 \end{equation}
where $N(t)$ is the Euclidean lapse function and $d\Omega_3^2$
is the metric on a unit round $S^3$.
The boundary conditions of regularity at the center are
$a(0)=0$, $\dot{a}(0)/N(0)=1$, and $\dot{\phi}(0)/N(0)=0$,
and the match to the boundary at $t=1$ gives
$a(1)=a_b$ and $\phi(1)=\phi_b$.

	If the scalar field potential is $[9/(16G^2)]V(\phi)$
(with the coefficient again chosen to simplify the formulas below
in terms of the rescaled potential $V(\phi)$), then the
Euclidean action of the history is
 \begin{eqnarray}
 I = -iS 
   &=& \int dt\left[{1\over 2N}(-a\dot{a}^2+a^3\dot{\phi}^2)
			+ {1\over 2}N(-a + a^3 V)\right]  \nonumber \\
   &=& -{1\over 2}\int dt\left[
   \tilde{N}^{-1}\tilde{G}_{AB}\dot{X}^A \dot{X}^B + \tilde{N}\right]
   = - \int d\tilde{s},
 \label{4}
 \end{eqnarray}
where
 \begin{equation}
 \tilde{N} = a\,(1-a^2 V)\,N \equiv e^{\alpha}\,(1-w)\,N,
 \label{5}
 \end{equation}
with
 \begin{equation}
 \alpha \equiv \ln{a},
 \label{6}
 \end{equation}
 \begin{equation}
 w \equiv a^2 V,
 \label{7}
 \end{equation}
and $d\tilde{s}$ is the infinitesimal proper distance
in the auxiliary two-metric
 \begin{eqnarray}
 d\tilde{s}^2 &=& \tilde{G}_{AB} dX^A dX^B
 	\nonumber \\
  &=& a^4 \left( 1-a^2 V \right) \left({da^2\over a^2} - d\phi^2 \right)
  \equiv e^{4\alpha} \left(1-w\right)\left( d\alpha^2 - d\phi^2 \right)
  	\nonumber \\
  &=& e^{2u+2v}\left(1-w\right) du dv
  = {1\over 4}\left(1-w\right) dX dY.
 \label{8}
 \end{eqnarray}
The auxiliary metric (\ref{8}) has null coordinates
 \begin{equation}
 u \equiv \alpha - \phi \equiv \ln{a} - \phi, \ \ 
 v \equiv \alpha + \phi \equiv \ln{a} + \phi,
 \label{9}
 \end{equation}
or alternate null coordinates
 \begin{equation}
 X \equiv e^{2u} \equiv e^{2\alpha-2\phi},\ \ 
 Y \equiv e^{2v} \equiv e^{2\alpha+2\phi}.
 \label{10}
 \end{equation}
 
	The zero-loop or classical histories are those that extremize
the Euclidean action (\ref{4}).  Define the rescaled proper Euclidean
time (or, for short, `Euclidean time')
 \begin{equation}
 \tau = \int_0^t N(t')dt' =
  \sqrt{3\pi\over 2G}({\rm proper\ radius\ or\ Euclidean\ `time'}),
 \label{11}
 \end{equation}
which is gauge invariant, invariant under reparametrizations
of the original time coordinate $t$ when the lapse function $N(t)$
is properly adjusted, though its value at the boundary
depends upon the particular history chosen.
Then extremizing the action with respect to $N(t)$
leads to the constraint equation
 \begin{equation}
 \left({da\over d\tau}\right)^2-a^2\left({d\phi\over d\tau}\right)^2
  = 1-a^2 V \equiv 1-w.
 \label{12}
 \end{equation}
Extremizing with respect to $\phi(t)$ leads to the scalar field equation
 \begin{equation}
 {d^2\phi\over d\tau^2}+{3\over a}{da\over d\tau}{d\phi\over d\tau}
  = {1\over 2}{dV\over d\phi},
 \label{13}
 \end{equation}
and extremizing with respect to $a(t)$
leads to the other field equation,
 \begin{equation}
 {1\over a}{d^2 a\over d\tau^2}+2\left({d\phi\over d\tau}\right)^2
 = - V,
 \label{14}
 \end{equation}
though either of these last two equations is redundant if one uses
the other along with the constraint equation. 

	Alternatively, zero-loop or classical histories
(extrema of the action)
are geodesics of the auxiliary two-metric (\ref{8}),
with this metric giving the proper distance along an extremum as
 \begin{equation}
 d\tilde{s} = -dI = \tilde{N}dt = a\,(1-a^2 V)\,d\tau.
 \label{15}
 \end{equation}
The Euclidean action $I$ is then simply the negative of the
proper distance from the center to the boundary along a geodesic
of the auxiliary metric.  If the center has $\tau=0$
and the boundary (where the wavefunction is being avaluated) has
 \begin{equation}
 \tau = \tau_b = \int_0^1 N(t) dt,
 \label{16}
 \end{equation}
then the Euclidean action for the classical history is
 \begin{equation}
 I = -\int_0^1\tilde{N}dt = -\int_0^{\tau_b}a\,(1-a^2 V)\,d\tau.
 \label{17}
 \end{equation}

\section{Real classical solutions for the FRW-scalar model}

	Therefore, a classical or extremal history
$(a(\tau),\phi(\tau))$
for the FRW-scalar model obeys the regularity conditions
$a=0$, $da/d\tau=1$, and $d\phi/d\tau=0$ at the center, $\tau=0$,
and so is uniquely determined,
for a given rescaled potential function $V(\phi)$,
by the value of $\phi$ at the center, $\phi_0 \equiv \phi(0)$,
and by the value of $\tau$ at the boundary, $\tau_b$,
where the wavefunction is being evaluated.
(One could alternatively have $da/d\tau=-1$ at the center,
but this would give
a negative value for the volume of the Euclidean geometry,
and the opposite sign of the Euclidean action, so I shall reject
this possibility.)

	Let us for simplicity restrict attention to analytic potentials
that for all real finite $\phi$ are real finite convex functions
that are bounded below by nonnegative values
(which can be considered to be nonnegative cosmological constants).
Since the action and extrema are invariant under replacing $\phi$
by $\phi' = -\phi$ if $V(\phi)$ is replaced by $V'(\phi') = V(\phi)$,
without loss of generality we can consider the case in which
the convex $V(\phi)$ is nondecreasing as $\phi$ is increased
from $\phi_0$, so $dV/d\phi \geq 0$ for $\phi \geq \phi_0$.
(We shall not need any further the condition that $V(\phi)$ be convex
and nonnegative for real $\phi$, but only that $V(\phi)$
be real, finite, positive, differentiable, and monotonically increasing
for $\phi > \phi_0$.)

	Then the scalar field equation (\ref{13}), rewritten as
 \begin{equation}
 {d\over d\tau}\left(a^3{d\phi\over d\tau}\right)
  = {1\over 2}a^3{dV\over d\phi},
 \label{18}
 \end{equation}
implies that $\phi$ cannot decrease with $\tau$.  Furthermore,
it can stay constant only if $\phi_0$ is at the minimum value
of $V(\phi)$, say $V_0$, in which case $\phi$ does stay at $\phi_0$,
and one gets $a = V_0^{-1/2}\sin{(V_0^{1/2}\tau)}$,
giving a 4-metric (\ref{3}) that is
part or all of a round 4-sphere of radius $(1.5\pi V_0/G)^{-1/2}$,
depending on whether $\tau_b$ is less than or equal to its
maximum value of $\pi V_0^{-1/2}$, where $a$ returns to zero.

	But if $dV/d\phi$ is positive at $\phi=\phi_0$ (and hence,
by assumption, remains positive for all larger $\phi$),
then Eq. (\ref{15}) has its right hand side become positive
as soon as $a(\tau)$ becomes positive, and so $a^3 d\phi/d\tau$
increases monotonically with real increasing $\tau$.
It thus follows that
$\phi$ and $V$ also increase monotonically with $\tau$.

	The field equation (\ref{14})
with real $a$, $\phi$, and $\tau$ and with positive $V$ implies
that $a$ is a concave function of $\tau$.
With $V$ not only positive but also increasing with $\tau$,
$a(\tau)$ necessarily reaches a finite maximum, say $a_m = a_m(\phi_0)$
at the Euclidean time $\tau_m = \tau_m(\phi_0)$,
with the functions $a_m(\phi_0)$ and $\tau_m(\phi_0)$
depending on the function $V(\phi)$,
and then $a$ returns to 0 at finite $\tau$, say $\tau_s=\tau_s(\phi_0)$,
with the function $\tau_s(\phi_0)$
also depending on the function $V(\phi)$.

	Since $a^3 d\phi/d\tau$ increases monotonically with $\tau$
under the assumptions above, $d\phi/d\tau$ becomes
infinite at $\tau=\tau_s$ as $a$ returns to zero,
giving a curvature singularity there,
and one can further show that $\phi$ goes to infinity there as well.

	Given a particular fixed rescaled potential $V(\phi)$
obeying the assumptions above for all real $\phi$
(real, finite, differentiable, and either monotonically increasing
with $\phi$ for all $\phi$ or else monotonically increasing
in both directions away from a single minimum),
we thus see that a choice of the two real parameters
$\phi_0$ and $\tau_b$ leads to a unique classical solution,
if $\tau_b < \tau_s(\phi_0)$,
and uniquely gives the boundary values $a_b = a(\phi_0,\tau_b)$
and $\phi_b = \phi(\phi_0,\tau_b)$, as well as the action
$I = I(\phi_0,\tau_b)$.

	Of course, we are really interested in evaluating
$\psi(a_b,\phi_b)$ by the zero-loop approximation
and hence want the action, for each of a suitable set of extrema,
as a function of $a_b$ and $\phi_b$ instead of as a function
of $\phi_0$ and $\tau_b$.  To do this, we need to solve
for values of the parameters $(\phi_0,\tau_b)$
that give the desired boundary values $(a_b,\phi_b)$.
Because the number of parameters matches the number
of boundary values, we expect a discrete set of solutions,
but the number of solutions may not be precisely one
for each $(a_b,\phi_b)$.

	For example, consider the case in which $V(\phi)$
is a slowly varying function of $\phi$.  In this case
the scalar field equation (\ref{13}) implies that $\phi$
does not change much during the evolution, so as a zeroth-order
approximation one can take $\phi \approx \phi_0$
and hence also $\phi_b \approx \phi_0$.
Then if one restricts to real values of $\tau_b$,
one sees that there are two solutions for $(\phi_0,\tau_b)$
given $(a_b,\phi_b)$ if $a_b < a_m(\phi_0) \approx a_m(\phi_b)$,
because the real classical Euclidean solution that starts
at the center, $a=0$, with real $\phi_0$ there,
has $a$ increasing from 0 to its maximum $a_m(\phi_0)$
at $\tau=\tau_m(\phi_0)$
and then decreasing to 0 again at $\tau=\tau_s(\phi_0)$,
so there are two solutions for $\tau_b$, one with $\tau_b < \tau_m$
in which $a$ crosses $a_b$ while increasing with $\tau$,
and the second with $\tau_b > \tau_m$
in which $a$ crosses $a_b$ while decreasing with $\tau$.

	On the other hand, if $a_b > a_m(\phi_0) \approx a_m(\phi_b)$,
there are no real solutions for $(\phi_0,\tau_b)$,
because all of the real solutions that match $\phi_b$
have maxima for $a$ that are smaller than $a_b$.

	One proposal would be simply to say that the zero-loop
approximation gives a wavefunction that is the sum of the $e^{-I}$'s
when there are one or more real classical solutions matching the real
boundary data that are the arguments of the wavefunction, and that the
zero-loop approximation gives zero when there are no real classical
solutions.  However, then one would simply get zero for most large
universes with nontrivial matter, such as $\psi_{\rm
0-loop}(a_b,\phi_b) = 0$ when $\phi_b$ is not at the minimum of the
potential that is zero there (or is at the minimum if the potential is
positive there) and when $a_b$ is sufficiently large (e.g., larger than
$V_0^{-1/2}$ if the rescaled potential has a positive minimum of
$V_0$).

	Furthermore, even for potentials allowing large classical
Euclidean universes, the action for each of them would be real, and so
the zero-loop approximate wavefunction would be purely real and not
have the oscillatory behavior apparently necessary to describe our
observations of an approximately Lorentzian universe.

	Therefore, it is not adequate to restrict
the zero-loop approximation to real classical Euclidean histories.

\section{Complex classical solutions for the FRW-scalar model}

	To get a potentially adequate zero-loop approximation,
we shall consider complex classical solutions (though still
with real boundary values that are the arguments of the wavefunction).
That is, we shall take the classical field equations
(\ref{12})-(\ref{14}) as complex analytic equations for
complex quantities $a$, $\phi$, $\tau$, and $V(\phi)$.
We shall assume that $V(\phi)$ is a complex analytic function
that for real values of $\phi$ has the properties assumed above
(real, finite, differentiable, and either monotonically increasing
with $\phi$ for all $\phi$ or else monotonically increasing
in both directions away from a single minimum).
The monotonicity property of $V(\phi)$
for real $\phi$ is not really important
but shall continue to be assumed here
to simplify some of the discussion.

	For a complex classical or extremal history
$(a(\tau),\phi(\tau))$ for the FRW-scalar model,
we shall continue to assume the regularity conditions
$a=0$, $da/d\tau=1$, and $d\phi/d\tau=0$ at the center, $\tau=0$,
as complex analytic equations that are the essential
input from the `one-boundary' proposal
when it is extended to allow complex solutions
in the zero-loop approximation.
Again the classical history is uniquely determined,
for a given rescaled potential function $V(\phi)$,
by the value of $\phi$ at the center, $\phi_0 \equiv \phi(0)$,
and by the value of $\tau$ at the boundary, $\tau_b$,
where the wavefunction is being evaluated.
The only difference is that both $\phi_0$ and $\tau_b$
may be complex, though we shall only be interested in solutions
that give real values for $a_b(\phi_0,\tau_b)$
and $\phi_b(\phi_0,\tau_b)$.

	For example, let us return to the case in which
$V(\phi)$ is slowly varying so that $\phi$ remains close
to $\phi_0$ throughout the classical history.
Then $\phi_b \approx \phi_0$, so for the desired real $\phi_b$,
we can take $\phi_0$ to be approximately real.
Now let us take $a_b$ to be much larger than $a_m(\phi_b)$,
so there is no real classical solution matching $(a_b,\phi_b)$
on the boundary.  However, we can find a complex classical
solution matching the boundary data in the following way:

	First, relative to the small variation of $V(\phi)$,
take the zeroth-order approximation that $\phi_0=\phi_b$.
Then the history is given by a contour in the complex $\tau$
plane from its value of 0 at the center to some complex value
$\tau_b$ at the boundary.  Consider the contour in which $\tau$
starts off real and increasing.  The rescaled $S^3$ size $a$
begins increasing as $\tau$, but as a concave function of real $\tau$,
it eventually reaches a maximum, $a_m(\phi_0)$,
at $\tau=\tau_m(\phi_0)$,
and then would decrease if $\tau$
continued to increase along its real axis.
Thus to lowest nontrivial order, $a(\tau)$ varies quadratically
with $\tau-\tau_m$ when this quantity is small,
with a negative coefficient.

	Therefore, after reaching $\tau_m$ on the real axis,
make a right-angled bend in the contour for $\tau$ so that now
$\tau-\tau_m$ takes on an imaginary value and $a$ continues to increase.
One can then follow a contour for $\tau$ so that $a$ stays real
and increases up to the desired boundary value $a_b$.

	If $V(\phi)$ were positive and precisely constant,
having no variation at all
and thus being equivalent to a cosmological constant,
the classical solution would have $\phi=\phi_0$ everywhere
and so would match the boundary condition if one chose $\phi_0=\phi_b$.
In this case the part of the contour with $\tau$ increasing along
the real axis from 0 to $\tau_m$ would give the geometry of a
Euclidean 4-hemisphere, and then the part of the contour
with $\tau-\tau_m$ changing in the imaginary direction
would give a Lorentzian deSitter universe expanding from
its time-symmetric throat at $\tau=\tau_m$ (to be discussed more below).

	If $V(\phi)$ is not precisely constant but has a slow variation
with $\phi$, and if $\phi_0$ is taken to be precisely real,
then along the part of the contour in which $\tau$ increases along
the real axis from 0 up to $\tau_m$,
$\phi$ will develop a small positive derivative, $d\phi/d\tau$,
and will increase slightly over its initial value of $\phi_0$
to become, say, $\phi_m$ at $\tau=\tau_m$, still real.
But when one turns the corner in the contour for $\tau$,
although $a$ had zero time derivative there and so could remain real,
$\phi$ has a small positive time derivative and so picks up
a small imaginary contribution as $\tau-\tau_m$ increases
(or decreases) in the imaginary direction, hence becoming slightly
complex at $\tau_b$ where $a$ matches the boundary value $a_b$.

	(The fact that $\phi$ becomes slightly complex
implies that the varying $V(\phi)$ also becomes slightly complex,
making the geometry slightly complex.  This would make $a(\tau)$
slightly complex, and hence never reaching the real boundary
value $a_b$, if one kept $\tau$ on a contour with
$\tau-\tau_m$ purely imaginary, though that contour
would keep the time-time part of the metric, proportional to $d\tau^2$,
purely real and negative or Lorentzian.  However, one can instead,
at least if $V(\phi)$ is slowly varying, distort the contour
for $\tau$ slightly to keep $a(\tau)$ precisely real,
and hence reaching $a_b$, but at the cost of making $d\tau^2$,
and hence the time-time part of the geometry, slightly complex.)

	However, just as one can compensate for a slightly
complex $a(\tau)$ along the simple-minded contour by distorting
the contour slightly, one can also compensate for a slightly
complex value of $\phi(\tau_b)$ by distorting the initial value
$\phi_0$ slightly into the complex, as several of us realized
independently but which Lyons
\cite{Lyons}
was the first to write down; see also
\cite{UJ,Page02}
for more recent work in this area.
To the lowest order in the slow variation of $V(\phi)$,
one can see how much error there is in $\phi(\tau_b)$
when one starts with the trial value $\phi_0=\phi_b$
and then evolves along the contour for $\tau$ that keeps $a(\tau)$
real and positive until one reaches $a=a_b$ at $\tau=\tau_b$,
and then correct that trial value of $\phi_0$ by the opposite
of that error.  One can take this first-corrected $\phi_0$
as a second trial value for $\phi_0$, follow a contour that goes from
$a=0$ to $a=a_b$ at a suitable $\tau=\tau_b$,
find the error in $\phi(\tau_b)$, and make a second correction.
In this way one can in principle iterate until one presumably finds,
to sufficient accuracy, the correct
complex $\phi_0$ that leads to $\phi=\phi_b$ at $a=a_b$.

	For a sufficiently rapidly varying $V(\phi)$,
this iteration procedure may not converge.
For example, there may be no contour that keeps $a$ real all
the way from 0 to $a_b$ and also allows $\phi=\phi_b$ there.
An example of this for an exponential potential will be given below.
However, there can be other complex solutions that can match
the boundary values $(a_b,\phi_b)$, even if there are none that
have $a$ stay purely real along some contour.  Of course, this
also highlights the possibility that even when the iterative procedure
above leads to a unique complex solution (up to complex conjugation;
see immediately below) matching the boundary values,
there may be other complex solutions
that also match the boundary values, and finding the criterion
for which to include in the zero-loop approximation may be problematic.

	The complex histories that lead to the desired real boundary
values will have complex Euclidean action $I$, with the imaginary
part depending on the boundary values that are the argument of the
wavefunction, so the contribution that they give to the
zero-loop wavefunction, $e^{-I}$, will be complex and have complex
oscillations as a function of the boundary values.
(This is true even if $V(\phi)$ is precisely constant and positive.)

	When $V(\phi)$ is an analytic function that is real
for real $\phi$, and when the boundary values $a_b$ and $\phi_b$
are real, as I am always assuming, then for any complex
initial data $(\phi_0,\tau_b)$ that leads to these real boundary values,
the complex conjugate data will lead to the same real boundary values
and thus also represent a history that matches the boundary values,
by the analyticity of the classical equations.
Therefore, complex classical histories always occur in pairs.
Since the actions of the two histories in each pair will also
be the complex conjugates of each other, when in the
zero-loop approximation one adds up the two complex conjugate
values of $e^{-I}$, one will always get a real sum.
No matter how many pairs of complex conjugate contributions one adds
(or whether one adds individual real contributions from
real classical solutions), one always gets a real wavefunction
(in this configuration representation), though it can be negative
and hence oscillate with the boundary values in a way that
the contributions of the purely real classical solutions could not.

	Of course, one would expect that this feature would persist
even if one multiplied the zero-loop contributions by prefactors
(say to incorporate one-loop determinants) or otherwise went beyond
the one-loop approximation, but this discussion shows that
a real oscillating wavefunction can arise simply out of the
zero-loop approximation if one allows the contributions from
complex conjugate pairs of classical solutions.

\section{FRW-scalar models with an exponential\\potential,
$V = e^{2\beta\phi}$}

	To illustrate some of these ideas quantitatively,
it is helpful to consider the case of an exponential potential,
 \begin{equation}
 V(\phi) = e^{2\beta\phi},
 \label{19}
 \end{equation}
where $\beta$ is a real parameter that characterizes
how fast the potential varies as a function of $\phi$.

	In terms of the quantities defined by Equations
(\ref{6}), (\ref{7}), (\ref{9}), and (\ref{10}), one then gets
 \begin{equation}
 w \equiv a^2 V = e^{2\alpha + 2\beta\phi}
   = e^{(1-\beta)u + (1+\beta)v} = X^{1-\beta\over 2}Y^{1+\beta\over 2}.
 \label{20}
 \end{equation}
The auxiliary two-metric (\ref{8}) is then
 \begin{eqnarray}
 d\tilde{s}^2
  &=& e^{4\alpha} \left(1-w\right)\left( d\alpha^2 - d\phi^2 \right)
  \nonumber \\
  &=& e^{2u+2v}\left(1-e^{(1-\beta)u + (1+\beta)v}\right) du dv
  \nonumber \\
  &=&{1\over 4}\left(1-X^{1-\beta\over 2}Y^{1+\beta\over 2}\right)dXdY.
 \label{21}
 \end{eqnarray}
This metric has a scaling symmetry,
exhibited by the homothetic Killing vector
 \begin{equation}
 {\bf K} = \left({1+\beta}\right){\partial\over\partial u}
          -\left({1-\beta}\right){\partial\over\partial v},
 \label{22}
 \end{equation}
whose action is to multiply $a^2 \equiv e^{2\alpha} \equiv e^{u+v}$
by a constant while keeping $w$ (or $\alpha+\beta\phi$) fixed,
thereby multiplying the metric (\ref{21})
by the square of this constant.
This is a symmetry that maps geodesics (which represent classical
solutions) onto geodesics, though multiplying the lengths of
the geodesics, and hence the actions of the solutions they represent,
by the same constant by which $a^2$ is multiplied.
It is this symmetry that allows one to reduce the two nontrivial
parameters of a generic two-metric to one nontrivial parameter
for the metric (\ref{21}) and to reduce the generic second-order
geodesic equation to a single first-order differential equation below.

	For the exponential potential (\ref{19}), the constraint
equation (\ref{12}) becomes
 \begin{equation}
 \left({d\alpha\over d\tau}\right)^2-\left({d\phi\over d\tau}\right)^2
  - e^{-2\alpha} + e^{2\beta\phi} = 0.
 \label{24}
 \end{equation}
The scalar field equation (\ref{13}) becomes
 \begin{equation}
 {d^2\phi\over d\tau^2} + 3{d\alpha\over d\tau}{d\phi\over d\tau}
  - \beta e^{2\beta\phi} = 0,
 \label{25}
 \end{equation}
and the field equation (\ref{14}) becomes
 \begin{equation}
 {d^2 \alpha\over d\tau^2} + \left({d\alpha\over d\tau}\right)^2
  + 2\left({d\phi\over d\tau}\right)^2 + e^{2\beta\phi} = 0.
 \label{26}
 \end{equation}
As before, these two second-order field equations are not independent
of each other if one uses the constraint (\ref{14}). 

	The symmetry action of the homothetic Killing vector upon
these classical field equations with an exponential potential
is to multiply the rescaled Euclidean time $\tau$, the rescaled $S^3$
size $a = e^{\alpha}$ and the inverse square root of the rescaled
potential, $V^{-1/2} = e^{-\beta\phi}$, all by the same constant.

	If we take out the scaling behavior represented
by the homothetic Killing vector of the auxiliary two-metric (\ref{21}),
we can represent the nontrivial behavior of the classical solutions
by using two scale-invariant quantities.  For one of them it is
convienient to use $w$ defined by Eq. (\ref{20}) above.
When $\beta \neq 0$, for the other it is convenient to use
 \begin{equation}
 u \equiv - {1\over\beta}{d\phi\over d\alpha}.
 \label{27}
 \end{equation}
I shall henceforth use this definition of $u$ rather than
the previous (different) use of $u$ for the null coordinate defined
in Eq. (\ref{9}).
The nontrivial behavior of a classical solution is then given
by the relation between $u$ and $w$, say $u(w)$ in a regime
where this function is single-valued.

	One can readily find that the relation between $u$,
$w$, and $\alpha$ is given by the following two equations:
 \begin{equation}
 {du\over d\alpha} = - \left({1-\beta^2 u^2\over 1-w}\right)
 			\left(2u+w-3uw\right),
 \label{28}
 \end{equation}
 \begin{equation}
 {dw\over d\alpha} = - 2 w (1-\beta^2 u).
 \label{29}
 \end{equation}
The relation with the Euclidean time $\tau$ is then given by
the constraint equation
 \begin{equation}
 \left({da\over d\tau}\right)^2 = {1-w \over 1-\beta^2 u^2}.
 \label{30}
 \end{equation}

	Now we see that we can divide Eq. (\ref{28}) by Eq. (\ref{29})
to get a single first-order differential equation relating
the two scale invariant quantities, $u$ and $w$:
 \begin{equation}
 {du\over dw} = - \left({1-\beta^2 u^2\over 1-\beta^2 u}\right)
 		        {2u+w-3uw \over 2w(1-w)}.
 \label{31}
 \end{equation}

	There is a one-parameter set of solutions of Eq. (\ref{31}),
say labeled by the initial condition $u(w_0)$ for some $w_0$.
In the FRW-scalar model, the center of the FRW geometry has,
if $V$ is finite there, $w=0$.  This is a singular point of
Eq. (\ref{31}), but one can readily show that the
regularity of the geometry there implies the `one-boundary' condition
 \begin{equation}
 u(w) = - {1\over 4}w + O(w^2).
 \label{32}
 \end{equation}
Analogous to what was done above, one should also take the positive
square root of the constraint equation (\ref{30}),
so that $da/d\tau = +1$ at the center and hence that the Euclidean
volume is positive near there for positive real $\tau$.

	Once one solves Eq. (\ref{31}) for $u(w)$ with the
`one-boundary' condition (\ref{32}), one can start at the boundary value
 \begin{equation}
 w_b = a_b^2 V(\phi_b)
 \label{33}
 \end{equation}
and choose a complex contour for $w$ to go to the center, $w=0$.
Along this contour, one can start at the boundary with the boundary
values $\alpha_b \equiv \ln{a_b}$ and $\phi_b$ and integrate
 \begin{equation}
 d\alpha = {dw \over 2w(1-\beta^2 u)}
 \label{34}
 \end{equation}
and
 \begin{equation}
 d\phi = -\beta u d\alpha = {-\beta u dw \over 2w(1-\beta^2 u)}
 \label{35}
 \end{equation}
to get $\alpha(w)$ and $\phi(w)$ along the contour.
One can also evaluate the action of this classical history as
 \begin{eqnarray}
 I(a_b,\phi_b) &=& -\int a^2\sqrt{1-w}\sqrt{d\alpha^2-d\phi^2}
 	\nonumber \\
   &=& - {1\over 2}\int_0^{w_b} dw e^{-2\beta\phi}\sqrt{1-w}
       {\sqrt{1-\beta^2 u^2}\over 1-\beta^2 u}.
 \label{36}
 \end{eqnarray}

 	However, one can also find that once $u_b \equiv u(w_b)$
is determined, the action is given algebraically by
 \begin{equation}
 I(a_b,\phi_b) = - {1\over 2} a_b^2 (1-u_b)
     \sqrt{1-w_b \over 1-\beta^2 u_b^2}.
 \label{37}
 \end{equation}
This may be derived by considering the fact that as a function
of the coordinates of the auxiliary two-metric (\ref{8}) for a general
FRW-scalar model, the action obeys the Hamiltonian-Jacobi equation
 \begin{equation}
 1=(\nabla I)^2 = G^{AB}I_{,A}I_{,B}
  ={e^{-4\alpha}\over 1-w}
   \left[\left({\partial I\over\partial\alpha}\right)^2
         -\left({\partial I\over\partial\phi}\right)^2\right].
 \label{38}
 \end{equation}

	When one restricts to the exponential potential so that
$w = e^{2\alpha + 2\beta\phi}$, one gets
the auxiliary two-metric (\ref{21}) with its homothetic Killing vector,
and one may look for solutions of the Hamilton-Jacobi equation
(\ref{38}) of the form
 \begin{equation}
 I = - {1\over 2} a^2 g(w),
 \label{39}
 \end{equation}
where $g(w)$ obeys the differential equation
 \begin{equation}
 \left(g+w{dg\over dw}\right)^2 - \beta^2 \left(w{dg\over dw}\right)^2
     = 1-w.
 \label{40}
 \end{equation}
One can then readily check that if $u(w)$ obeys the differential
equation (\ref{31}), then
 \begin{equation}
 g(w) = (1-u)\sqrt{1-w \over 1-\beta^2 u^2}.
 \label{41}
 \end{equation}
obeys the differential equation (\ref{40}).

	In terms of a classical solution of the field equations
(\ref{24})-(\ref{26}), say written as $e^{\alpha} = a(\tau)$
and $\phi = \phi(\tau)$, the action may alternatively be written as
 \begin{equation}
 I = - {1\over 2} a^2 \left({da\over d\tau}
     + {a\over\beta}{d\phi\over d\tau}\right).
 \label{41b}
 \end{equation}
 
	So far I have written as if there were a unique $u(w)$,
obeying the differential equation (\ref{31}) with the
`one-boundary' regularity condition (\ref{32}) at the center,
for each point of the complex $w$-plane.
Indeed it is true that if one starts with the regularity
condition (\ref{32}) at the center and integrates (\ref{31})
outward along some contour in the complex $w$-plane that avoids
the singular points of that differential equation,
one gets a unique answer for $u(w)$ along that contour,
and the analyticity of the differential equation
(away from its singular points) guarantees that the result
for $u(w)$ is independent of deformations of the contour
that do not cross any of the singular points.
However, because the differential equation (\ref{31})
does have singular points (at $w=0$, $w=1$, and $w=\infty$,
and at $u=1/\beta^2$ if $\beta^2$ is neither 0 nor 1;
in the latter case the zero in the denominator of the right hand side
of Eq. (\ref{31}) is canceled by the zero in the numerator),
the result for $u(w)$ generically depends upon
the topology of the contour relative to the singular points.
Thus one gets different Riemann sheets in which $u(w)$
has different values.
In particular, when one goes to the boundary value $w_b$
and evaluates $u_b = u(w_b)$ and then the action $I(a_b,\phi_b)$
by Eq. (\ref{37}), the action will, generically,
depend upon the topology of contour in the complex $w$-plane
from the center at $w=0$ to the boundary at $w=w_b$.

	For example, if one contour leads to a complex value
of $u(w_b)$ (even though we are restricting to $w_b$ real),
there will be a complex conjugate contour that leads to the
complex conjugate value of $u(w_b)$ and hence to the
complex conjugate value of the action $I(a_b,\phi_b)$.
However, generically there will be far more than a single pair
of topologically inequivalent contours leading from $w=0$
to $w=w_b$.  Typically there will be an infinite number
of such pairs, corresponding to winding around the various
singularities arbitrarily many times.  Because there are more
than one singularity, one could presumably wind around one singularity
an arbitrary (integer) number of times, then around another
an arbitrary number of times, then around another, and so on
ad infinitum, which would give an uncountably infinite number
of infinite contours.  However, if one restricted to finite
contours, there would be merely a countable infinity of them.

	One might also allow both signs for the square root
in Eq. (\ref{37}) for the action, but I would argue that
one should not do that for a given contour in the complex $w$-plane.
For a contour that stays at small real $w$
in going from 0 to a small real $w_b$ with a real Euclidean geometry,
one would want the four-volume to be positive and hence for
the action to be negative, $I \approx -a_b^2/2$.
This requires that one take the positive square root when $1-w$ and
$1-\beta^2 u^2$ are both real and near unity, thus determining which
branch of the square root to choose in the part of a contour
when it just leaves the center, $w=0$.  For any contour emerging
from the center (so long as it does not pass through the singular point
$w=1$ and also avoids $\beta^2 u^2 = 1$), one can follow the sign of
the square root continuously as one goes from $w=0$ to $w=w_b$
to get a unique answer there.

	However, this consideration does show that if one deforms
a contour to wrap once around a point with $\beta^2 u^2 = 1$,
this continuity requirement on the branch choice for the square
root will cause it to switch sign relative to the undeformed
contour that did not wrap around the $\beta^2 u^2 = 1$ point.
Thus the value of the action will depend not only on the topology
of the contour relative to the singular points
$w=0$, $w=1$, $w=\infty$, and, generically, $u=1/\beta^2$
of the differential equation (\ref{31}),
but its sign will also depend upon its topology
relative to the points $u=1/\beta$ and $u=-1/\beta$.

\subsection{$\beta=0$ deSitter example, $V =$ const.}

	There are two values of $\beta^2$ for which one can
explicitly solve the differential equation (\ref{31})
and get the classical solutions and their action,
$\beta^2 = 0$ and $\beta^2 = 1$.
(Reversing the sign of $\beta$ is equivalent to reversing the
sign of the scalar field and has no effect
on the differential equation (\ref{31}) or on the action.)

	When $\beta = 0$, the potential is independent of $\phi$.
For the exponential potential $V(\phi) = e^{2\beta\phi}$ given
by Eq. (\ref{19}), this would give $V=1$, but one can easily
generalize the result to any constant $V$.

	It is convenient to define
 \begin{equation}
 x \equiv \sqrt{1-w} = \sqrt{1-Va^2}
 \label{41c}
 \end{equation}
as a useful replacement for $w$ in certain equations.
I shall choose the positive sign at the center, so there $x=1$.
However, for real $a > 1/\sqrt{V}$, $x$ will be purely imaginary
and can have either sign, depending on which side of the singular
point $w=1$ the contour is taken, and whether it wraps around that
point an even or odd number of times.

	The `one-boundary' solution for constant $V$ is
 \begin{equation}
 \phi = \phi_b,
 \label{41d}
 \end{equation}
 \begin{equation}
 a = {1 \over \sqrt{V}}\sin{(\sqrt{V}\tau)},
 \label{42}
 \end{equation}
 \begin{equation}
 w = \sin^2{(\sqrt{V}\tau)},
 \label{43}
 \end{equation}
 \begin{equation}
 x = \cos{(\sqrt{V}\tau)}.
 \label{44}
 \end{equation}

	For real $\tau$ ranging up to $\pi/\sqrt{V}$,
one gets part of a round Euclidean $S^4$ with equatorial
$S^3$ of rescaled radius
 \begin{equation}
 a_m = a(\tau_m) = {1 \over \sqrt{V}}
 \label{45}
 \end{equation}
at
 \begin{equation}
 \tau_m = {\pi\over 2\sqrt{V}}.
 \label{46}
 \end{equation}
If $\tau$ ranges all the way from 0 to $2\tau_m$,
one gets a complete Euclidean $S^4$.

	On this $S^4$, define the latitudinal angle
 \begin{equation}
 \theta = \sqrt{V}(\tau_m - \tau),
 \label{47}
 \end{equation}
so
 \begin{equation}
 a = a_m\cos{\theta},\ x=\sqrt{1-w}=\sin{\theta}.
 \label{48}
 \end{equation}

	Now to get to real $a > a_m$, as discussed above
for a general slowly varying positive potential,
have $\tau$ go along its real axis from $\tau=0$ to $\tau=\tau_m$
but then make a right-angled bend in the complex $\tau$-plane,
so that $\tau-\tau_m$ becomes henceforth imaginary.
In particular, analytically continue $\theta$ to $\theta = \pm i \psi$
with $\psi$ real to get
 \begin{equation}
 a = a_m\cosh{\psi},\ x=\sqrt{1-w}=\pm i\sinh{\psi}.
 \label{49}
 \end{equation}
Then the four-metric (\ref{3}) becomes
 \begin{equation}
 ds^2 = \left({2G\over 3\pi V}\right)
 	\left(-d\psi^2 + \cosh^2{\psi}d\Omega_3^2\right),
 \label{50}
 \end{equation}
which is the real Lorentzian deSitter spacetime.

	Although strictly speaking $u$ as defined by Eq. (\ref{27})
is not well defined when $\beta=0$, as both the numerator and
the denominator are zero, it is well defined when one starts with
$\beta \neq 0$ and then takes the limit of $\beta$ going to zero.
In particular its differential equation (\ref{31}), and its
`one-boundary' regularity condition (\ref{32}) at the center,
are both well-defined when $\beta=0$ and lead to the solution
 \begin{equation}
 u = {1\over 3}\left(1-{2\over 1-w+\sqrt{1-w}}\right)
   = {(x-1)(x+2)\over 3x(x+1)},
 \label{51}
 \end{equation}
one example of an equation that is simpler in terms of
$x \equiv\sqrt{1-w}$ than in terms of $w \equiv 1-x^2$.

	Unlike the case of generic $\beta$, for $\beta=0$
there is only the two-fold ambiguity in $u(w)$, depending
on the choice of the sign of the square root
of $1-w$ for $x=\sqrt{1-w}$.

	When one uses this $u(w)$ or $u(x)$ in Eq. (\ref{37})
for the Euclidean action, one gets
 \begin{equation}
 I = {x_b^3-1\over 3V} = {(1-w_b)^{3/2}-1\over 3V}
   = {\sin^3{\theta_b}-1\over 3V} = {\pm i \sinh{\psi_b}-1\over 3V}.
 \label{52}
 \end{equation}
Thus the Euclidean action is real (and negative) for $w_b < 1$
(or $a_b < 1/\sqrt{V}$) but is complex (but still with a negative
real part) for $w_b > 1$ (or $a_b > 1/\sqrt{V}$).

	In the special case of $\beta=0$, there also seems to be
no possibility of different topologies of the contour relative
to the points that are at $\beta^2 u^2 = 1$ for nonzero $\beta$,
so in this special case there does not seem to be the possibility
to reverse the overall sign of the action, assuming that one
always starts the contour from $w=0$ and the choice of the sign
of the square root in Eq. (\ref{37}) or (\ref{41}) so that the
four-volume starts off becoming positive.

	For $a_b \geq 1/\sqrt{V}$, one gets
 \begin{equation}
 |e^{-I}|^2 = e^{+{2\over 3V}},
 \label{53}
 \end{equation}
the famous Hartle-Hawking enhancement of the relative probabilities
that is greater for smaller positive $V$
\cite{HNP}.

\subsection{$\beta=1$ example, $V = e^{2\phi}$}

	This special exponential potential has been discussed by
\cite{GHM},
who give the solutions in terms of different variables,
but here I shall follow my previous notation to show how they
fit into my scheme for a general exponential potential.

	For $\beta=1$, the auxiliary two-metric (\ref{21})
becomes the flat metric
 \begin{equation}
 d\tilde{s}^2 = {1\over 4}(1-Y)dXdY = {1\over 4}(1-w)dXdw = dXdZ,
 \label{54}
 \end{equation}
where now $X = a^2/V = a^4/w$ and $Y = a^2 V = w$,
and I have defined $Z = (2w-w^2)/8$ to get the metric
into the explicitly flat form with null coordinates $X$ and $Z$.

	One can show that only for an exponential potential,
and then only for $\beta = \pm 1$,
is the general auxiliary two-metric (\ref{8}) flat
for the general FRW-scalar model.

	The generic geodesic of the flat auxiliary two-metric
(\ref{54}) has the form
 \begin{equation}
 X = CZ+D,
 \label{55}
 \end{equation}
where $C$ and $D$ are arbitrary constants.
The `one-boundary' condition of regularity at the center, $a=0$,
implies that the exponential potential $V$ must be finite and nonzero
there, so both $X$ and $Z$ should vanish there, giving $D=0$
\cite{GHM,Unruh}.
Then
 \begin{equation}
 X \equiv {a^2\over V} \equiv {a^4\over w} \equiv {w\over V^2}
   = CZ \equiv {C\over 8}(2w-w^2) = {C\over 8}(2a^2 V - a^4 V^2),
 \label{56}
 \end{equation}
so
 \begin{equation}
 C = {8a_b^4\over w_b^2(2-w_b)}={8e^{-4\phi_b}\over 2-a_b^2 e^{2\phi}},
 \label{57}
 \end{equation}
 \begin{equation}
 V^2 = {8\over C(2-w)} = {V_b^2(2-a_b^2V_b)\over 2-w},
 \label{58}
 \end{equation}
 \begin{eqnarray}
 a^4 &=& {C\over 8}w^2(2-w) = a_b^4{w^2(2-w)\over w_b^2(2-w_b)}
 	\nonumber\\ 
     &=& {1\over V^2}\left(2-{8\over CV^2}\right)^2
      =  \left({a_b^2 V_b^3-2V_b^2\over V^3} + {2\over V}\right)^2.
 \label{59}
 \end{eqnarray}
 
 	One can further show that the differential equation (\ref{31}),
with the `one-boundary' regularity condition (\ref{32}) at the center,
leads to the solution
 \begin{equation}
 u = {-w\over 4-3w} = {1\over 3}\left(1-{4\over 4-3w}\right)
   = {x^2-1\over 3x^2+1}.
 \label{60}
 \end{equation}
In this case (only) there is no branch-cut ambiguity at all in $u(w)$,
though there is a two-fold ambiguity in the action when it is complex,
as there must be, since for any `one-boundary' regular classical
solution with real boundary values $(a_b,\phi_b)$,
its complex conjugate, with a complex conjugate action,
must also be a regular classical solution by the analyticity
properties of the field equations and the regularity conditions.

	As the negative of the geodesic path length in the
the flat auxiliary two-metric (\ref{54}), the Euclidean action
can easily be evaluated to be
 \begin{equation}
 I = -\sqrt{X_b Z_b} = -{1\over 2}a_b^2\sqrt{1-{1\over 2}a_b^2 V_b}.
 \label{61}
 \end{equation} 

	For $0 < w_b < 2$, one gets $C > 0$
and a real Euclidean solution with real (negative) action.
Unlike the case of $\beta=0$, for which the action approached
the negative value $-1/(3V)$ at the boundary of the Euclidean
region, there at $w_b = 1$ or $x_b = 0$, here at the boundary
of the Euclidean region (now at $w_b = 2$ or $x_b = \pm i$)
the Euclidean action approaches zero.  Along any real Euclidean
classical solution, the action becomes negative and decreases as $w$
increases from 0 to 1, and then the action increases back to zero
as $w$ further increases from 1 to 2.
(It is interesting that the turning point for the action, at $w=1$,
does not coincide with the turning point for $a$,
which increases from zero to a maximum
as $w$ increases from 0 to 4/3 and then decreases back to zero
as $w$ further increases from 4/3 to 2.)

	For $2 < w_b$, one gets $C < 0$,
and the geodesic in the $(X,w)$ coordinates has, if one follows
a contour of real $w$, $X$ starting off zero at zero $w$ and then
goes negative, decreasing as $w$ increases from 0 to 1 and then
increasing back to zero again as $w$ increases from 1 to 2.
In this region Eqs. (\ref{58}) and (\ref{59}) show that
$V^2$ and $a^4$ and are both negative, so the geometry is complex.
Although the complex $a$ returns to 0 at $w=2$, the purely
imaginary $V$ goes to infinity there, giving a singularity
in the metric,
though one that could be avoided by choosing the contour
in the complex $w$-plane to avoid the point $w=2$.

	It is interesting that this singular point in the geometry
at $w=2$ is not a singular point, when $\beta^2=1$,
of the differential equation (\ref{31}),
which then has singular points only at $w=0$, $w=1$, and $w=\infty$.
Conversely, $w=0$ is not a singularity of the geometry when
the regularity condition (\ref{32}) is imposed, $w=1$
is just a coordinate singularity in the flat auxiliary metric (\ref{54})
that also gives no physical singularity, and $w=\infty$ simply gives
a universe that has expanded to infinite size and zero potential
(if $w_b > 2$
and if $w$ is taken to $\infty$ along the positive real axis).

	If one continues to follow a contour of real $w$ for $w_b > 2$,
when $w > 2$ one gets a precisely real Lorentzian four-metric,
 \begin{equation}
 ds^2 = \left({2G\over 3\pi}\right) (-C)^{1/2}
   {\sin{2\eta}\over\cos^3{2\eta}}\left(-d\eta^2 + d\Omega_3^2\right),
 \label{62}
 \end{equation}
for conformal time $\eta$ in the range $0 < \eta < \pi/4$,
and defined, up to sign and up to shifts by arbitrary multiples of
$\pi/2$, by
 \begin{equation}
 w = 2 \sec^2{2\eta}.
 \label{63}
 \end{equation}
During the expansion of this universe from a big bang at $\eta=0$
to infinite size at $\eta=\pi/4$, the rescaled potential is given by
 \begin{equation}
 V = e^{\pm 2\phi} = 2 (-C)^{-1/2} \cot{2\eta}
 \label{64}
 \end{equation}
and thus drops from an infinite value at the big bang to
a value of zero at infinite expansion.

	Unlike the case in which the potential $V(\phi)$
is slowly varying, the exponential potential $V=e^{2\beta\phi}$
with $\beta^2=1$ does not allow one to follow a contour that keeps
$a$ real as one goes from the regular center to an asymptotic
region where the geometry is asymptotically real and Lorentzian
(in this case precisely real and Lorentzian).
This one can see from Eq. (\ref{59}) for $a^4(w)$ with $C < 0$.
If one starts from $w=0$ along a contour that keeps $a$ real
(initially having $w$ depart from 0 in the purely imaginary direction),
then if one follows the contour with $a$ real, one will for large $w$
have it asymptotically become a large (and growing) real number
multiplied by one of the two complex cube roots of unity,
rather than having it join the real axis where the metric
is real and Euclidean.

	From purely looking at this Lorentzian big-bang solution,
which is singular at $a=0$,
one would hardly guess that it also obeys the `one-boundary'
regularity condition at $a=0$,
but from the analysis above we see that it does,
having both a singular big bang singularity with $a=0$ and infinite $V$
at $w=2$ or $\eta=0$, and also having a regular center with $a=0$
and finite (though complex) $V = V_b \sqrt{1-w_b/2}$ at $w=0$ or
$\eta = \pm i\infty$.

	Thus we see that for a constant potential ($\beta=0$),
an analytic continuation of the `one-boundary' condition of
regularity at $a=0$ leads to a precisely real and Lorentzian
geometry, the deSitter four-metric (\ref{50}), which is nonsingular
everywhere and has no big bang or big crunch.
For the exponential potential with $\beta^2 = 1$, $V = e^{\pm 2\phi}$
for the rescaled scalar field $\phi$, one also gets that
an analytic continuation of the `one-boundary' condition of
regularity at $a=0$ leads to a precisely real and Lorentzian
geometry, the four-metric (\ref{62}), but this time it is a singular
metric with a big bang in which both the potential energy and
the kinetic energy densities of the scalar field are infinite.

\subsection{Generic $\beta$ in the exponential potential
$V = e^{2\beta\phi}$}

	For $\beta^2$ different from 0 or 1, the singularities
of the differential equation (\ref{31}) generically lead
to nontrivial branch cuts, except for the one at $w=1$,
which is typically accompanied by having $u$ pass through $\pm 1/\beta$
so that all quantities behave regularly there.
The singular point at $w=0$ is also generically accompanied
by having $u$ pass through $\pm 1/\beta$
(except at the beginning of a contour for which one imposes
the `one-boundary' regularity condition to avoid a singularity there),
but if indeed $u=\pm 1/\beta$ at $w=0$, this represents
a curvature singularity, and going around it in different ways
in the complex plane can lead to different results for $u(w)$
(being on different Riemann sheets).
The singular point at $w=\infty$ cannot be accompanied by
$u=\pm 1/\beta$ but is accompanied by $u=1/3$ and represents
a universe that has expanded to infinite size.
It is also a branch-cut singularity, so that one gets different
results for $u(w)$ depending on the topology of the contour
relative to that singularity, as well at to the one at $w=0$.

	For $\beta^2 < 3$, a solution for $u(w)$ can be analytically
continued to give an asymptotically real Lorentzian solution
with large real $w$ and nearly real $a$, $\phi$, and rescaled
proper Lorentzian time $t_L = -i\tau$, with
 \begin{equation}
 u \sim {1\over 3}
     + {2\over 9}\left({9-\beta^2\over 3+\beta^2}\right){1\over w}
     + O\left({1\over w^3}\right)
     + C_1(\beta^2)\left(-{1\over w}\right)
                  ^{\left({9-\beta^2\over 6-2\beta^2}\right)}
		       \left[1+O\left({1\over w}\right)\right],
 \label{65}
 \end{equation}
where $C_1(\beta^2)$ is some function of $\beta^2$.
One can then use this to find that the Euclidean action
at large $|w|$ asymptotically goes as
 \begin{equation}
 I\!\sim\!{1\over \sqrt{9\!-\!\beta^2} V}\!
       \left[\!(\!-w)^{3\over 2}
             \!+\! {3\over 2}A(\!-w)^{1\over 2}
	     \!-\! A C_2(\beta^2)(\!-w)^{\!-{\beta^2\over 3-\beta^2}}
 	     \!-\! {3\over 8} A^2 (\!-w)^{\!-{1\over 2}}\!\right],
 \label{66}
 \end{equation}
where
 \begin{equation}
 A \equiv \left({\sqrt{9-\beta^2}\over 3+\beta^2}\right).
 \label{66b}
 \end{equation}
For simplicity I have dropped the subscripts $b$ on the boundary values,
but what is written here as $V$ and $w$ should actually be $V_b$
and $w_b$.
Here $C_2(\beta^2)$ is another function of $\beta^2$, derivable
from $C_1(\beta^2)$.  $C_2(0) = C_2(1) = 1$, and according
to a preliminary approximate equation I have obtained for $C_2(\beta^2)$
\cite{Page02},
apparently it lies between approximately 0.94 and 1
for $0 \leq \beta^2 \leq 1$.

	Eq. (\ref{66}) gives the behavior of the action
``near'' the singular point $w=\infty$ (i.e., for $|1/w| \ll 1$),
with the Euclidean action
being purely real (and positive) if one takes the contour
in the complex $w$-plane to run along the negative real axis
(which of course is a rather unphysical region with $w\equiv a^2 V<0$).
Then to get to positive $w$, one can follow $1/w$ around 0 in its
complex plane from its negative axis to its positive axis.
Let us define the integer $n$ to be zero if one takes $1/w$
counterclockwise half way around its zero,
from the negative real axis to
the positive real axis for $1/w$ in the sense that goes below 0
in the complex plane.  Then let nonzero $n$ represent the number
of excess times the contour is taken counterclockwise around $1/w = 0$.
E.g., if the contour were then taken once clockwise around
after reaching the positive real axis from the counterclockwise
half-rotation, so that the net effect would be a clockwise
half-rotation, passing once above 0 in going from the negative
axis to the positive axis for $1/w$, then $n=-1$.

	As a result of this, after reaching the positive $w$
axis with some integer $n$ representing the integer part
of the winding number, one can make the replacement
 \begin{equation}
 \left(-{1\over w}\right) = e^{(2n+1)\pi i}\left({1\over w}\right)
 \label{67}
 \end{equation}
in the asymptotic formula (\ref{66}) for the action, giving
 \begin{eqnarray}
 I\sim\!\! &\!\!\!-\!\!\!&\!\!{1\over 3 V}\!
 		\left({\sqrt{9\!-\!\beta^2}\over 3\!+\!\beta^2}\right)
  		C_2(\beta^2)w^{-{\beta^2\over 3-\beta^2}}
   \left[\cos{(2n\!+\!1)\pi\beta^2\over 3-\beta^2}
         \!+\!i\sin{(2n\!+\!1)\pi\beta^2\over 3-\beta^2} \right]
	   \nonumber \\
  &\!\!\!-\!\!\!&\!\!{i(-1)^n\over \sqrt{9-\beta^2}\, V} \left[w^{3/2}
                  -{3\over 2} A w^{1/2}
		  +{3\over 8} A^2 w^{-1/2}
		  +O(w^{-3/2})\right].
 \label{68}
 \end{eqnarray}
 
 	This formula represents only a tiny class of possible
contours for $w$ in going from the regular center at $w=0$
to the asymptotically Lorentzian regime of a large universe
with $w \gg 1$, with only the single integer $n$,
representing how many times around the singularity at $w=\infty$
the contour wraps.  Thus this class of contours does not
include the possibility of wrapping around the singularity at $w=0$
an arbitrary number of times between each possible
wrapping around the singularity at $w=\infty$.
That much larger class, though still not including the possibility
of going around the singularity at $u=1/\beta^2$ (which seems to give
merely a two-fold square-root ambiguity
and so may have less effect than the other singularities),
would be parametrized by an arbitrary sequence of integers and signs
(with each integer in the sequence being the number of times
wrapping around the singularity at $w=0$
between each time of wrapping once around the singularity at $w=\infty$,
and with each sign being the sign of the wrapping around
the singularity at $w=\infty$).
However, here we are not wrapping around the singularity at $w=0$
at all, so the small subclass being considered is characterized by
the single integer $n$.

	The first term in the Euclidean action above is the real
part of the Euclidean action, $I_R$, which gives the magnitude
of the zero-loop approximation to the wavefunction if only that
history contributes, $|e^{-I}|^2 = e^{-2I_R}$.  If $n=0$ or $n=-1$,
the two simplest contours, then $I_R$ is negative for $\beta^2 < 1$
(assuming that $C_2(\beta^2)$ remains positive)
and vanishes for $\beta^2 = 1$, this last fact being consistent
with the exact solution given above for $\beta^2 = 1$.
Presumably these simplest contours then make $I_R$ positive
when $\beta^2$ is increased beyond 1.  {\it If} $C_2(\beta^2)$
remains positive for all $\beta^3$ up to 3, then the simplest contours
would make $I_R$ oscillate an infinite number of times as $\beta^2$
is increased to 3, the maximum value it can have and still lead to
an approximately Lorentzian universe
that can expand to arbitrarily large size.
For example, if the asymptotic formula (\ref{68}) were accurate for
the real part of the action, $I_R$,
that quantity would oscillate and pass through zero at
 \begin{equation}
 \beta^2 = {6N-3 \over 2N+1}
 \label{69}
 \end{equation}
for each positive integer $N$ (unrelated to $n$, which is being set
equal to 0 or -1 to give the simplest possible contours),
assuming that $C_2(\beta^2)$ does not pass through infinity
at any of those points to prevent $I_R$ from passing through zero.

	If the winding number $n$ can take on arbitrary integer values,
then for values of $\beta^2$ other than those given by Eq. (\ref{69}),
$I_R$ can take on both positive and negative values.
This highlights the question of what classical histories
are to be included if the zero-loop approximation is supposed
to be a reasonable approximation to the `one-boundary' wavefunction.

	In models in which $Re\sqrt{g}\,$, the real part of the proper
four-volume per coordinate four-volume, has a fixed sign,
Halliwell and Hartle
\cite{HalHar}
proposed that one should include only classical histories in which
the sign is positive.  The most conservative interpretation of this
in models such as those being considered here, in which the sign of
$Re\sqrt{g}$ can vary along a contour, is that one should choose
the square root of this metric determinant $g$
to give a real value in the part of contour near the center,
which is the same as my choice of the sign of the square root
in Eq. (\ref{41}) to make my $g(w)$ (not the determinant of the metric)
positive for small $w$ in the part of the contour
near its beginning at the center, $w=0$.

	But one might also interpret the Halliwell-Hartle proposal
as requiring that, for a monotonically increasing real time
coordinate $t$ along the complex contour for $w$, one have
$Re\sqrt{g} \propto Re(a^3 d\tau/dt)$
be positive along the entire contour.  It remains to be seen whether
this can always be satisfied for the models considered above.
(It does have the ugliness of depending on the details of the contour
even within the same topological class of how it winds around the
various singularities, whereas one would prefer simply a restriction
to a particular topological class.)  But if it can, one might conjecture
that it might be satisfied only for the contours given above
with $n=1$ and $n=-1$ (which are a complex-conjugate pair).

	Alternatively, one might simply postulate that for
the FRW-scalar model with an exponential potential,
for large $w_b$ one should just use the simplest
single complex-conjugate pair of contours, those given above with
$n=0$ and $n=-1$.  Both of these have the same real part, $I_R$,
and opposite imaginary parts, say $I_I$ for $n=0$,
so $n=0$ gives $I=I_R + i I_I$ and $n=-1$ gives the complex conjugate
$I' = I^{\star} = I_R - i I_I$.
(In general the action for some $n$ is the complex conjugate of the
action with a new integer $n' = -n-1$, e.g., $n'=-1$ for $n=0$.)
Then if we take the zero-loop approximation to be given
by those two contributions, we get
\begin{equation}
\psi_{\rm 0-loop}(a_b,\phi_b) = e^{-I} + e^{-I'} = 2e^{-I_R}\cos{I_I}.
\label{70}
\end{equation}

	However, it would be highly desirable to have some
physically motivated principles for selecting a suitable
set of complex classical solutions for the zero-loop approximation.

\section{Summary}

\begin{itemize}

\item  The Hartle-Hawking `one-boundary' proposal
is one of the first attempts to say which wavefunction,
out of all those satisfying the dynamical equations,
corectly describes our universe.

\item  There are severe problems doing the path integral it calls for.

\item  At the zero-loop level it makes a number of
remarkable predictions (large nearly flat Lorentzian universe,
second law of thermodynamics, etc.)

\item  However, even here there are generically an infinite
number of complex extrema to choose from,
and it is not quite clear how to do this properly.
Certainly one important goal of this approach to quantum cosmology
would be to give a specification of which extrema to use,
and then of course one would like to compare the results
with as many observations as possible.

\end{itemize}

\section{Acknowledgments}

	I am deeply indebted to Stephen Hawking for being
an excellent mentor for me as a graduate student, postdoc,
and professor.  His courage in tackling physical challenges
has been personally inspiring, and his courage in tackling
physics challenges has been academically inspiring.
He has opened up a whole new approach to physics and cosmology
in seeking to find the quantum state of the universe,
and I am certain that the path he has begun will become
an integral part of our understanding of the world.

	For the particular work summarized here,
I am not only indebted to Hawking's work but have also
benefited from recent discussions with Jonathan Halliwell
and Bill Unruh.  This work was financially supported in part
by the Natural Sciences and Engineering Council of Canada.

%\printindex
\end{document}